\documentclass[12pt]{article}
\usepackage{amsmath}
\usepackage{graphicx,psfrag,epsf}
\usepackage{enumerate}
\usepackage{natbib}
\usepackage{url} 
\usepackage{bm}
\usepackage{amsfonts}

\usepackage{multirow}
\usepackage{comment}
\usepackage[utf8]{inputenc}
\usepackage[english]{babel}
\usepackage{amsthm}
\theoremstyle{plain}
\newtheorem{theorem}{Theorem}[section]
\newtheorem{property}{Property}
\newtheorem{lemma}[theorem]{Lemma}
\newcommand{\code}[1]{\texttt{#1}}
\newcommand{\blind}{1}
\usepackage[symbol]{footmisc}

\begin{document}

\def\spacingset#1{\renewcommand{\baselinestretch}%
{#1}\small\normalsize} \spacingset{1}


\if1\blind
{
  \title{\bf Statistical Inference of Auto-correlated Eigenvalues with Applications to Diffusion Tensor Imaging}
  \author{Zhou Lan\\\
    Center for Outcomes Research and Evaluation \\
    Yale School of Medicine}
  \maketitle
} \fi

\if0\blind
{
  \bigskip
  \bigskip
  \bigskip
  \begin{center}
    {\LARGE\bf Statistical Inference of Auto-correlated Eigenvalues with Applications to Diffusion Tensor Imaging}
\end{center}
  \medskip
} \fi

\bigskip
\begin{abstract}
Diffusion tensor imaging (DTI) is a prevalent neuroimaging tool in analyzing the anatomical structure. The distinguishing feature of DTI is that the voxel-wise variable is a $3\times3$ positive definite matrix other than a scalar, describing the diffusion process at the voxel. Recently, several statistical methods have been proposed to analyze the DTI data. This paper focuses on the statistical inference of eigenvalues of DTI because it provides more transparent clinical interpretations. However, the statistical inference of eigenvalues is statistically challenging because few treat these responses as random eigenvalues. In our paper, we rely on the distribution of the Wishart matrix's eigenvalues to model the random eigenvalues. A hierarchical model which captures the eigenvalues' randomness and spatial auto-correlation is proposed to infer the local covariate effects. The Monte-Carlo Expectation-Maximization algorithm is implemented for parameter estimation. Both simulation studies and application to IXI data-set are used to demonstrate our proposal. The results show that our proposal is more proper in analyzing auto-correlated random eigenvalues compared to alternatives. 
\end{abstract}

\noindent%
\vfill

\newpage
\spacingset{1.45} 
\section{Introduction}
\begin{figure}[ht!]
    \centering
    \includegraphics[width=0.3\textwidth]{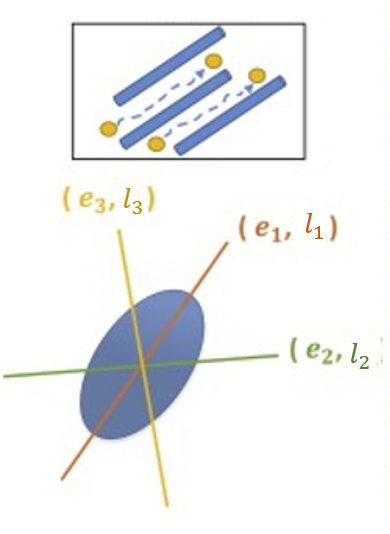}
    \caption{The figure gives the graphical illustration of the anatomical structure and the corresponding diffusion process. The upper panel gives the tissue to be measured. The lower panel gives the corresponding diffusion ellipsoid.}
    \label{fig:DTI}
\end{figure}

Diffusion tensor imaging (DTI) is a prevalent neuroimaging tool in analyzing the anatomical structure of a tissue. The rich use of DTI in brain science triggers many meaningful interdisciplinary studies. Due to the nature of interdisciplinarity, providing a tutorial of DTI, which is intuitive to both statisticians and clinicians, is key in interdisciplinary researches of DTI. Based on Figure \ref{fig:DTI} which is modified from figures in \citep[][Figure 1.6]{zhou2010statistical}, we give a short tutorial of DTI as follows. DTI is a neuroimage measuring the white matter tracts within a tissue (e.g., human brain) voxel-by-voxel. Let the structure at the upper panel of Figure \ref{fig:DTI} be the anatomical structure measured at a voxel. The movement of water molecules in different environments is measured by giving their respective diffusion ellipsoids at the lower panel of Figure \ref{fig:DTI}. The eigenvectors are the diffusion directions $[\bm{e}_1\ \bm{e}_2\ \bm{e}_3]$ and the squared roots of the eigenvalues $[l_1\ l_2\ l_3]$ are the corresponding semi-axis lengths \citep[][Section 1.2.3]{zhou2010statistical}. The corresponding diffusion tensor $D=[\bm{e}_1\ \bm{e}_2\ \bm{e}_3]\text{diag}(l_1,l_2,l_3)[\bm{e}_1\ \bm{e}_2\ \bm{e}_3]^T$ describes the anatomical structure at the voxel. Combining the diffusion tensors, which are positive definite over all the voxels, we have an image whose voxel-wise variables are positive definite diffusion tensors, describing the anatomical structure of a tissue.

The use of DTI is extensive (e.g., fiber tracking, clinical studies, etc.). Our paper focuses on understanding how the subject's covariates (e.g., age, gender, disease status) drive the brain's anatomical structures in terms of DTI. Many brain scientists raised this scientific objective. The relevant researches are now divided into two paths. The statisticians are interested in developing methodologies that handle the statistical randomness of the whole diffusion tensor. There are many representative works \citep[e.g.,][]{schwartzman2008inference,zhu2009intrinsic,yuan2012local,lee2017inference,spatialwishart}. The clinicians are more interested in using convenient summary scalar to tell the changes of diffusion tensor quickly and easily. For example, the fractional anisotropy, denoted as ${\displaystyle {\text{FA}}={\sqrt {\frac {1}{2}}}{\frac {\sqrt {(l _{1}-l _{2})^{2}+(l _{2}-l _{3})^{2}+(l _{3}-l _{1})^{2}}}{\sqrt {l _{1}^{2}+l _{2}^{2}+l _{3}^{2}}}},}$ is a frequently used summary scalar. Via implementing basic statistical tools, i.e., t-test, simple linear regression with these quantities as data, the researchers can easily tell the changes of the tissue statistically \citep[e.g.,][]{ma2009diffusion,lane2010diffusion}. 

In light of Figure \ref{fig:DTI}, we propose that analyzing the DTI's eigenvalues and eigenvectors separately can be another promising research direction. For statisticians, this path does not preclude any valuable available data information. For clinicians, the scientific explanations in terms of the statistical uncertainties of eigenvalues and/or eigenvectors are more transparent than the other ways, e.g., coefficients based on Cholesky decomposition \citep{zhu2009intrinsic,spatialwishart} or local polynomial regression \citep{yuan2012local}. In our paper, we focus on the statistical analysis of eigenvalues because the variations of eigenvalues are primarily interested by the clinicians given the current clinical reports \citep[e.g.,][]{ma2009diffusion,lane2010diffusion}. In this paper, the analysis of eigenvectors is left as future works.

Compared to other statistical models regarding the eigenvalues of DTI \citep[e.g.,][]{zhu2010multivariate,zhu2011fadtts,jin20173}, our paper may be a very first paper discussing the statistical inference of random eigenvalues. It is an important fact that the eigenvalues are stochastically ordered, and thus the probability density function of the eigenvalues must ensure the inequality of the eigenvalues stochastically. This inequality makes many classic models such as multivariate Gaussian processes not be applied to this problem straightforwardly. In this paper, we use the density function of a Wishart distributed matrix's eigenvalues as the basic probability distribution, referred to as \textit{distribution of Wishart's eigenvalues}. The distribution of Wishart's eigenvalues ensures the inequality of the eigenvalues stochastically.

\begin{figure}[ht!]
    \centering
    \includegraphics[width=0.8\textwidth]{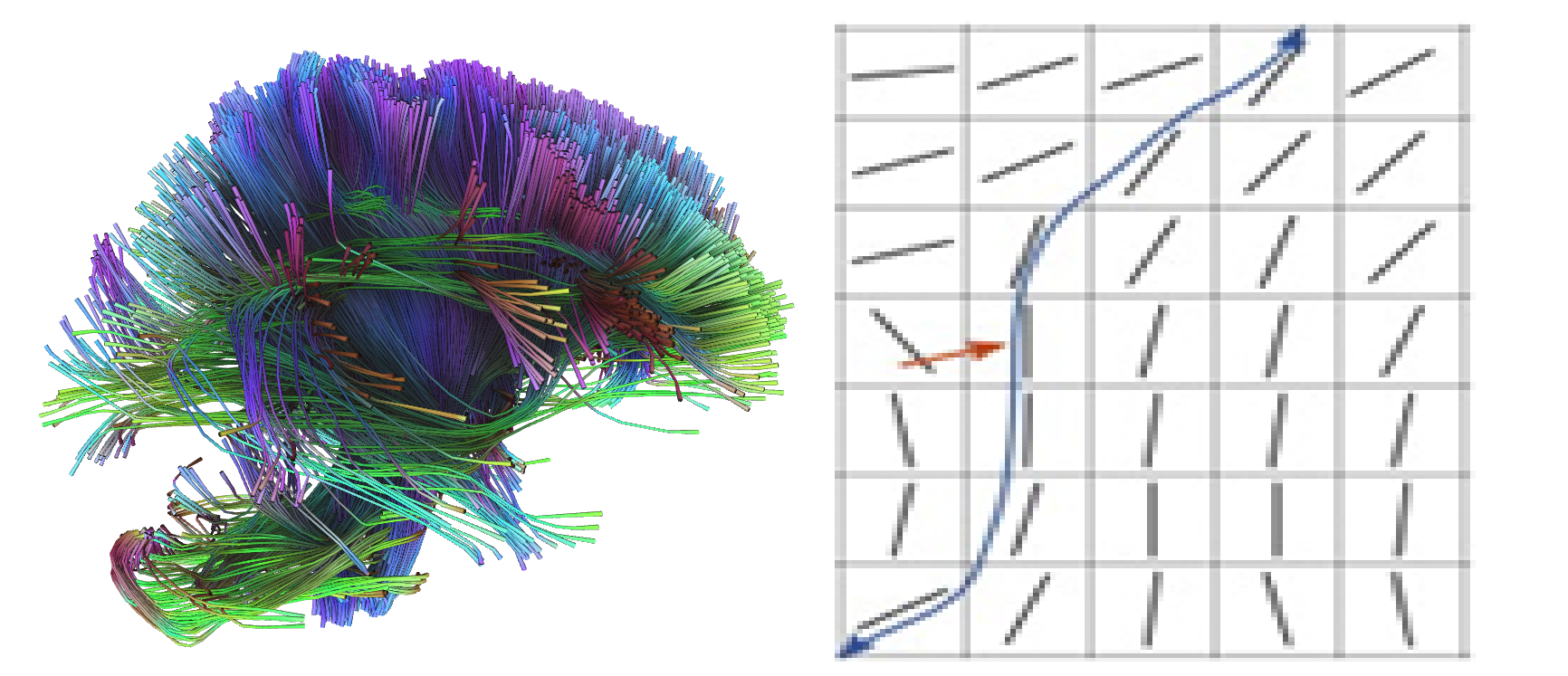}
    \caption{The left panel gives a human brain in terms of fiber tracts. The right panel gives how fiber tracts are constructed. The principal eigenvalues of diffusion tensors are given in each voxel. The voxels with similar directions are considered as a fiber tract, e.g., the voxels linked by the blue curve.}
    \label{fig:fiber}
\end{figure}

Following many neuroimaging studies, the spatial dependence of the eigenvalues among voxels is encouraged to be incorporated. In the previous analysis \citep[e.g.,][]{zhu2010multivariate,zhu2011fadtts,jin20173}, the spatial dependence can be easily achieved via the spatial Gaussian process model. Many fiber tracking reports \citep[e.g.,][]{wong2016fiber} assume that the diffusion tensors are smooth along with a fiber tract (see Figure \ref{fig:fiber}) whose diffusion tensors have similar directions. In light of this assumption, we assume that the eigenvalues are auto-correlated. The model based on auto-correlated eigenvalues captures the spatial correlation and produces an analytic expression of the density function of the spatially correlated eigenvalues \citep{james1964distributions}. 

 Combining all the above, we construct a hierarchical spatial model for the random auto-correlated eigenvalues. This model is both statistically and scientifically proper regarding the applications to diffusion tensor imaging. Given the hierarchical structure, we implement the Monte-Carlo Expectation-Maximization (EM) algorithm \citep[][Section 2.1]{levine2001implementations} for the parameter estimation and implement parametric bootstrapping for uncertainties quantification.
 
Both simulation studies and real data applications are used to demonstrate our proposed method. We compare our proposal to a non-linear regression model treating the data as normally distributed. The simulation studies indicate that our model gives a more accurate parameter point and confidence interval estimation. The real data analysis based on the IXI data-set (\url{https://brain-development.org/ixi-dataset/}) provided by Center for the Developing Brain, Imperial College London, an open data-set made available under the Creative Commons CC BY-SA 3.0 license, shows more clinically meaningful results after treating the data as random eigenvalues.
 
 In the rest of the paper, we first give the hierarchical spatial model of random auto-correlated eigenvalues in Section \ref{sec:model}. The estimation of the proposed model is provided in Section \ref{sec:est}. We give simulation studies and real data analysis in Section \ref{sec:sim} and Section \ref{sec:app}, respectively. Finally, we give a conclusion in Section \ref{sec:con}. The theoretical details in our paper are in Appendix \ref{sec:proofs}. The codes for implementing our method are summarized in Appendix \ref{sec:codes}.

\section{Model Setup}
\label{sec:model}
In this section, we give the model construction step-by-step. First, we provide the basic model setup in Section \ref{sec:basic}, where the probability distribution of the eigenvalues and the covariate effect specification are introduced. Under the basic model, we further give a latent layer of eigenvalues that induce spatial correlation and provide the details in Section \ref{sec:latent}. To wrap up, we provide a model summary and introduce its application to the statistical inference of eigenvalues of DTI in Section \ref{sec:model_summary}.

Our model is based on the the parameterized Wishart distribution, and the \textit{parameterized} Wishart distribution is slightly different from the Wishart distribution introduced in the classic textbooks. We give the construction of the parameterized Wishart distribution as follows. $\bm{Z}_j$ independently follows a mean-zero normal distribution with the covariance matrix $\bm{\Sigma}$ over $j\in\{1, ..., U\}$ , denoted as $\bm{Z}_j\stackrel{ind}{\sim}\mathcal{N}(\bm{0},\bm{\Sigma})$. $\bm{W}=\sum_{j=1}^U\bm{Z}_j\bm{Z}_j^T/U$ follows a parameterized Wishart distribution with the mean matrix $\bm{\Sigma}$ and the degrees of freedom $U$, denoted as $\bm{W}\sim\mathcal{W}_p(\bm{\Sigma},U)$. The density function of $\bm{W}$ is given in Appendix \ref{sec:density}.

\subsection{Basic Model Setup}
\label{sec:basic}
Let $\bm{D}_{it}$ be a $3\times 3$ positive definite matrix-random variable following a parameterized Wishart distribution with the mean matrix $\bm{\Sigma}_{it}$ and the degrees of freedom $m$, denoted as
\begin{equation}
    \bm{D}_{it}\sim\mathcal{W}_3(\bm{\Sigma}_{it},m).
\end{equation}
$\bm{D}_{it}$ can be treated as the diffusion tensor at the voxel $t\in\{1,2,...,T\}$ of the subject $i\in\{1,2,...,I\}$. The eigenvalues of $\bm{D}_{it}$ are $[L_{it}^{(1)} ,L_{it}^{(2)}, L_{it}^{(3)}]$ with the stochastic inequality $L_{it}^{(1)}>L_{it}^{(2)}>L_{it}^{(3)}$. Besides the degrees of freedom $m$, the density function of $[L_{it}^{(1)} ,L_{it}^{(2)}, L_{it}^{(3)}]$ only depends on the eigenvalues of $\bm{\Sigma}_{it}$ which are $\lambda_{it}^{(1)}> \lambda_{it}^{(2)} >\lambda_{it}^{(3)}$ \citep[][Theorem 3.2.18]{muirhead2009aspects}. Therefore, we say that $[L_{it}^{(1)} ,L_{it}^{(2)}, L_{it}^{(3)}]$ follow a distribution of Wishart's eigenvalues with the degrees freedom $m$ and the \textit{population} eigenvalues $\lambda_{it}^{(1)}> \lambda_{it}^{(2)} >\lambda_{it}^{(3)}$, denoted as
\begin{equation}
\label{eq:bbb}
    [L_{it}^{(1)} ,L_{it}^{(2)}, L_{it}^{(3)}]\sim\mathcal{E}([\lambda_{it}^{(1)}, \lambda_{it}^{(2)} ,\lambda_{it}^{(3)}],m)
\end{equation}
Given \citet[][Equation 3]{lawley1956tests}, the mean of the eigenvalue is expressed as $\mathbb{E} L_{it}^{(k)}=\lambda_{it}^{(k)}+\mathcal{O}(m^{-1/2})$, where $\mathcal{O}(m^{-1/2})$ is a quantity with order $m^{-1/2}$. Since $\lambda_{it}^{(k)}$ is the so-called \textit{population} eigenvalue of the distribution (the eigenvalue of the mean matrix $\bm{\Sigma}_{it}$), we point out that the expression $\mathbb{E} L_{it}^{(k)}=\lambda_{it}^{(k)}+\mathcal{O}(m^{-1/2})$ may not be consistent with our intuition. The value of the degrees of freedom $m$ also leads to the small variance of the eigenvalues. Therefore, $\lambda_{it}^{(k)}$ can be approximately treated as the mean of $L_{it}^{(k)}$ when $m$ is large enough, denoted as $\mathbb{E} L_{it}^{(k)}\simeq\lambda_{it}^{(k)}$, although its formal definition is the so-called \textit{population} eigenvalue.

Let $\bm{X}_{i}$ be a $1\times D$ matrix containing the covariate information (e.g., age, gender) of the subject $i$. We rescale the original value of each covariate $X_{id}$ into the range in $[0,1]$. We further specify that $\mathbb{E} L_{it}^{(k)}\simeq e^{\beta_{0t}^{(k)}+\bm{X}_i\bm{\beta}_{t}^{(k)}}=\lambda_{it}^{(k)}$ where $\bm{\beta}_{t}^{(k)}=[{\beta}_{1t}^{(k)},...,{\beta}_{Dt}^{(k)}]^T$, where ${\beta}_{dt}^{(k)}$ is the coefficient for the covariate $d$ of the $k$-th eigenvalue located at the voxel $t$ and ${\beta}_{0t}^{(k)}$ is the corresponding intercept. This specification ensures that the \textit{population} eigenvalues are positive, and the coefficients represent the linear effects on the logarithm of the mean of the random eigenvalue, denoted as $\log\mathbb{E} L_{it}^{(k)}\simeq\beta_{0t}^{(k)}+\bm{X}_i\bm{\beta}_{t}^{(k)}$. We also put the constraints ${\beta}_{dt}^{(1)}>{\beta}_{dt}^{(2)}>{\beta}_{dt}^{(3)}$ for $d=0,1,2,...,D$ to ensure the inequality of the \textit{population} eigenvalues. Combining all above, the means of the random eigenvalue responses (i.e., $\mathbb{E} L_{it}^{(k)}$) are determined by the covariate information.

\subsection{Latent Eigenvalues Setup}
\label{sec:latent}
In Section \ref{sec:basic}, we give basic model construction for the random eigenvalue responses. In this section, we further induce the spatial dependence among the random eigenvalues by giving the latent eigenvalues setup. The spatial model to be built is expected to have the following three properties: 
\begin{property}
The eigenvalues $[L_{it}^{(1)} ,L_{it}^{(2)}, L_{it}^{(3)}]$ are independent and identically distributed (\textit{i.i.d}) over subjects $i\in\{1,...,N\}$.
\end{property}
\begin{property}
Given a subject $i$, $[L_{it}^{(1)} ,L_{it}^{(2)}, L_{it}^{(3)}]$ are spatially correlated over the voxels $t\in\{1,...,T\}$.
\end{property}
\begin{property}
The approximation $\mathbb{E} L_{it}^{(k)}\simeq e^{\beta_{0t}^{(k)}+\bm{X}_i\bm{\beta}_{t}^{(k)}}$ given in the basic model setup is preserved after giving the latent eigenvalues setup.
\end{property}

To achieve these goals, we decompose $\lambda_{it}^{(k)}$ as $\lambda_{it}^{(k)}=\xi_{it}^{(k)}e^{\beta_{0t}^{(k)}+\bm{X}_i\bm{\beta}_{t}^{(k)}}$, where $\bm{\beta}_{t}^{(k)}=[{\beta}_{1t}^{(k)},...,{\beta}_{Dt}^{(k)}]^T$. The additional terms $\xi_{it}^{(1)}>\xi_{it}^{(2)}>\xi_{it}^{(3)}$ are random eigenvalues follow a so-called \textit{spatial process of Wishart's eigenvalues}, inducing the spatial correlation. As a prerequisite, we first introduce the spatial Wishart process \citep{gelfand2004nonstationary,spatialwishart} which is a means to induce the correlation among Wishart matrices. Let $\bm{Z}_t$ be a mean-zero Gaussian process with the covariance as $\text{cov}(\bm{Z}_t,\bm{Z}_s)=\rho_{ts}\times\bm{\Sigma}$, where $\rho_{ts}\in(0,1)$ is the correlation, and $\bm{\Sigma}$ is the cross covariance matrix of the Gaussian process. In the rest of the paper, we simply give $\bm{\Sigma}=\bm{I}$, where $\bm{I}$ is an identity matrix. If we have \textit{i.i.d} realizations of this process, denoted as $\bm{Z}_{jt}$, where $j$ is the subscript for realizations and $t$ is the subscript for the Gaussian process, then $\bm{U}_t=\sum_{j=1}^M\bm{Z}_{jt}\bm{Z}_{jt}^T/M$ follows a so-called spatial Wishart process, denoted as $\{\bm{U}_t\}_{t\in \mathbb{T}}\sim\mathcal{SWP}(M,\rho_{ts}\otimes\bm{I})$. Subsequently, the eigenvalues of $\bm{U}_t$ which are $\xi_{t}^{(1)}>\xi_{t}^{(2)}> \xi_{t}^{(3)}$ follow a spatial process of Wishart's eigenvalues, denoted as 
\begin{equation}
\label{eq:SEP}
{[\xi_{t}^{(1)}, \xi_{t}^{(2)}, \xi_{t}^{(3)}]\}_{t\in \mathbb{T}}\sim\mathcal{SEP}(M,\rho_{ts}\otimes[1,1,1])}. 
\end{equation}
Let $[\xi_{it}^{(1)}>\xi_{it}^{(2)}> \xi_{it}^{(3)}]$ be the $i$-th \textit{i.i.d.} realization of $\{[\xi_{t}^{(1)}, \xi_{t}^{(2)}, \xi_{t}^{(3)}]\}_{t\in \mathbb{T}}\sim\mathcal{SEP}(M,\rho_{ts}\otimes[1,1,1])$ at the voxel $t$. To wrap-up, our model at this stage is
\begin{equation}
\label{eq:full0}
\begin{aligned}
& [L_{it}^{(1)} ,L_{it}^{(2)}, L_{it}^{(3)}]|[\xi_{it}^{(1)},\xi_{it}^{(2)}, \xi_{it}^{(3)}]\sim\mathcal{E}([\lambda_{it}^{(1)}, \lambda_{it}^{(2)} ,\lambda_{it}^{(3)}],m),\\
       &\lambda_{it}^{(k)}=\xi_{it}^{(k)}e^{\beta_{0t}^{(k)}+\bm{X}_i\bm{\beta}_{t}^{(k)}},\\
       &\{[\xi_{it}^{(1)}, \xi_{it}^{(2)}, \xi_{it}^{(3)}]\}_{t\in \mathbb{T}}\sim\mathcal{SEP}(M,\rho_{ts}\otimes[1,1,1]).\\
\end{aligned}
\end{equation}
Next, we show this model satisfies the three properties we stated at the beginning of this subsection.

\begin{figure}[ht!]
    \centering
    \includegraphics[width=0.6\textwidth]{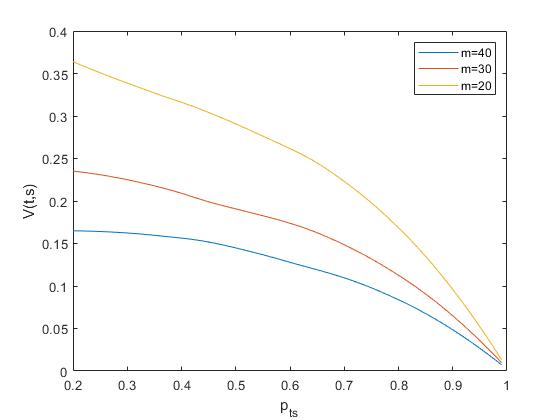}
    \caption{The function $\mathcal{V}(t,s)$ is visualized via Monte-Carlo approximation. The x-axis is for the values of $\rho_{ts}$. The y-axis is for the values of $\mathcal{V}(t,s)$. The orange, red, and blue curves are for $m=20,30,40$, respectively.}
    \label{fig:variogram}
\end{figure}

Since the latent eigenvalues $[\xi_{it}^{(1)}>\xi_{it}^{(2)}> \xi_{it}^{(3)}]$ are \textit{i.i.d.} over $i\in\{1,...,N\}$, Property 1 that the eigenvalues $[L_{it}^{(1)} ,L_{it}^{(2)}, L_{it}^{(3)}]$ are \textit{i.i.d} over subjects $i\in\{1,...,N\}$ is satisfied. We measure the spatial dependence of this process via the expected squared Frobenius norm of the term $\text{diag}([\xi_{it}^{(1)}, \xi_{it}^{(2)}, \xi_{it}^{(3)}])-\text{diag}([\xi_{s}^{(1)}, \xi_{s}^{(2)}, \xi_{s}^{(3)}])$, denoted as
\begin{equation}
\label{eq:variogram}
    \mathcal{V}(t,s)=\mathbb{E}||\text{diag}([\xi_{it}^{(1)}, \xi_{it}^{(2)}, \xi_{it}^{(3)}])-\text{diag}([\xi_{is}^{(1)}, \xi_{is}^{(2)}, \xi_{is}^{(3)}])||_F^2.
\end{equation}
The expected squared Frobenius norm $\mathcal{V}(t,s)$ can be treated as a variogram \citep{cressie2015statistics,spatialwishart}. We use Monte-Carlo approximation to visualize $\mathcal{V}(t,s)$ (see Figure \ref{fig:variogram}) since it has no analytic expression. The term $\mathcal{V}(t,s)$ is a decreasing function of $\rho_{ts}$. A larger value of the degrees of freedom $m$ leads to a larger value of $\mathcal{V}(t,s)$, indicating lager local variations. We conclude that the spatial dependence replies on the correlation parameter $\rho_{ts}$ of the latent Gaussian processes. Thus, Property 2 that $[L_{it}^{(1)} ,L_{it}^{(2)}, L_{it}^{(3)}]$ are spatially correlated over the voxels $t$, is satisfied. Furthermore, conditional on $\xi_{it}^{(k)}$, the expectation of $L_{it}^{(k)}$ is $\mathbb{E}_{L_{it}^{(k)}|\xi_{it}^{(k)}}L_{it}^{(k)}\simeq\xi_{it}^{(k)}e^{\beta_{0t}^{(k)}+\bm{X}_i\bm{\beta}_{t}^{(k)}}$. Locally, $[\xi_{it}^{(1)}, \xi_{it}^{(2)}, \xi_{it}^{(3)}]$ follows $\mathcal{E}([1,1,1],M)$ with $\mathbb{E}\xi_{it}^{(k)}\simeq 1$ as $M$ is large. Thus,  Property 3 that the approximation $\mathbb{E} L_{it}^{(k)}\simeq e^{\beta_{0t}^{(k)}+\bm{X}_i\bm{\beta}_{t}^{(k)}}$ is still preserved after $\xi_{it}^{(k)}$ is marginalized out. 

Next, we give the specification of the correlation $\rho_{ts}$. In light of many interdisciplinary works on fiber tracking \citet[e.g.,][Section 4]{wong2016fiber}, the spatial dependence can be assumed to be auto-correlated along with a fiber tract, and there is no spatial dependence across fibers (see Figure \ref{fig:fiber}). To accompany with this feature, we no longer use the subscript $t$ to denote the voxels but use the subscript $g_t$ to denote the $t$-th voxel of the $g$-th fiber. Given a fiber $g$, $[g_1, ..., g_t, ..., g_{T_g}]$ are the $1$-st, ..., $t$-th, ..., $T_g$-th diffusion tensor ordered from the first diffusion tensor of the fiber to the last diffusion tensor of the tensor, respectively. To accompany with the auto-correlation of the voxels along a fiber, we give that $\rho_{g_t,g'_{s}}=p^{|t-s|}$ for $g=g'$, where $p$ is the correlation parameter controlling the spatial dependence. To accompany with that there is no spatial dependence across fibers, we give that $\rho_{g_t,g'_{s}}=0$ for $g\not=g'$. A by-product of this spatial auto-correlation is to produce an analytic expression of the density function of the spatially correlated eigenvalues \citep{james1964distributions}.

\subsection{Model Summary and Diffusion Tensor Imaging}
\label{sec:model_summary}
To this end, our proposed model of random auto-correlated eigenvalues is fully expressed as
\begin{equation}
\label{eq:full1}
\begin{aligned}
&{\textbf{Basic Level:}}\left\{
\begin{aligned}
& [L_{i,g_t}^{(1)} ,L_{i,g_t}^{(2)}, L_{i,g_t}^{(3)}]|[\xi_{i,g_t}^{(1)},\xi_{i,g_t}^{(2)}, \xi_{i,g_t}^{(3)}]\sim\mathcal{E}([\lambda_{i,g_t}^{(1)}, \lambda_{i,g_t}^{(2)} ,\lambda_{i,g_t}^{(3)}],m);\\
       &\lambda_{i,g_t}^{(k)}=\xi_{i,g_t}^{(k)}e^{\beta_{0,g_t}^{(k)}+\bm{X}_i\bm{\beta}_{g_t}^{(k)}};\\
       &\bm{\beta}_{g_t}^{(k)}=[{\beta}_{1,g_t}^{(k)},...,{\beta}_{D,g_t}^{(k)}]^T;\quad {\beta}_{d,g_t}^{(1)}>{\beta}_{d,g_t}^{(2)}>{\beta}_{d,g_t}^{(3)} \\
       \end{aligned}
       \right.\\
             &{\textbf{Latent Level:}}\left\{
\begin{aligned}
       &\{[\xi_{i,g_t}^{(1)}, \xi_{i,g_t}^{(2)}, \xi_{i,g_t}^{(3)}]\}_{t\in \mathbb{T}}\sim\mathcal{SEP}(M,\rho_{g_t,g'_s}\otimes[1,1,1]);\\
            & \rho_{g_t,g'_s}=\left\{
\begin{aligned}
p^{|t-s|} & , & g=g' \\
0 & , & g\not=g'
\end{aligned}
\right.,
\end{aligned}\right.
\end{aligned}
\end{equation}
for $i=1,2,...,N;\ g=1,2,...,G;\ t=1,2,...,T_g$. 

This proposed model can be used for the statistical inference of eigenvalues of DTI. The DTI data are collected for subjects $i=1,2,...,N$. For each subject $i$, the eigenvalues of the voxel $t\in \{1,2,...,T_g\}$ located at the fiber $g\in\{1,2,...,G\}$ are treated as responses, i.e., $[L_{i,g_t}^{(1)} ,L_{i,g_t}^{(2)}, L_{i,g_t}^{(3)}]$. Conditional on $[\xi_{i,g_t}^{(1)},\xi_{i,g_t}^{(2)}, \xi_{i,g_t}^{(3)}]$, the eigenvalues $[L_{i,g_t}^{(1)} ,L_{i,g_t}^{(2)}, L_{i,g_t}^{(3)}]$ follow a distribution of Wishart's eigenvalues, i.e., $\mathcal{E}([\lambda_{i,g_t}^{(1)}, \lambda_{i,g_t}^{(2)} ,\lambda_{i,g_t}^{(3)}],m)$. The value of the degrees of freedom $m$ controls the local variations of the eigenvalues: a larger value of $m$ leads to smaller local variations. The term $e^{\beta_{0,g_t}^{(k)}+\bm{X}_i\bm{\beta}_{g_t}^{(k)}}$ dominates the mean of the eigenvalue $L_{i,g_t}^{(k)}$ when $m$ and $M$ are large enough. The coefficient ${\beta}_{d,g_t}^{(k)}$ captures the averaging effect of the covariate $d$ on the mean of $L_{i,g_t}^{(k)}$. 

To capture the spatial correlations, the latent variables $\{[\xi_{i,g_t}^{(1)}, \xi_{i,g_t}^{(2)}, \xi_{i,g_t}^{(3)}]\}_{t\in \mathbb{T}}$ follow the spatial process of Wishart's eigenvalues $\mathcal{SEP}(M,\rho_{g_t,g'_s}\otimes[1,1,1])$, where $\rho_{g_t,g'_s}=\left\{
\begin{aligned}
p^{|t-s|} & , & g=g' \\
0 & , & g\not=g'
\end{aligned}
\right.$. The value of the degrees of freedom $M$ controls the local variations of the latent eigenvalues. No correlation is assumed across fibers. The parameter $p$ controls the correlation among voxels within a fiber tract: a larger value of $p$ leads to strong correlations. The information of the fibers and their connected voxels us treated as \textit{priori}. These pieces of the known \textit{priori} information can be obtained from a human brain diffusion tensor template \citep[e.g.,][]{peng2009development}.

\section{Estimation}
\label{sec:est}
In this section, we introduce the statistical estimation of Model \ref{eq:full1}. We use $\bm{l}_{i,g_t}=[l_{i,g_t}^{(1)}, l_{i,g_t}^{(2)}, l_{i,g_t}^{(3)}]$ to denote the observed eigenvalues of the random eigenvalues $\bm{L}_{i,g_t}=[L_{i,g_t}^{(1)}, L_{i,g_t}^{(2)}, L_{i,g_t}^{(3)}]$. We also give $\bm{\xi}_{i,g_t}=[\xi_{i,g_t}^{(1)}, \xi_{i,g_t}^{(2)}, \xi_{i,g_t}^{(3)}]$, $\bm{L}=\{\bm{L}_{i,g_t}: i=1,2,...,N; g=1,2,...,G; t=1,2,...,T_g\}$, and $\bm{\xi}=\{\bm{\xi}_{i,g_t}: i=1,2,...,N; g=1,2,...,G; t=1,2,...,T_g\}$. The parameters to be estimated are $\bm{\theta}=\{M,m, p,\bm{\beta}^{(1)},\bm{\beta}^{(2)},\bm{\beta}^{(3)}\}$ where $\bm{\beta}^{(k)}=\{\bm{B}_{g_t}^{(k)}=[{\beta}_{0,g_t}^{(k)},{\bm{\beta}_{g_t}^{(k)}}^T]^T:g=1,2,...,G; t=1,2,...,T_g\}$.

First, we aim to give the explicit expression of the log-likelihood of the data $\bm{l}$ and the latent variables $\bm{\xi}$, in order to proceed with the next parameter estimation. The relevant proofs and derivations in this section are provided in Appendix \ref{sec:derivations}.

 The log-likelihood of the data $\bm{l}$ and the latent variables $\bm{\xi}$ is expressed as
\begin{equation}
\label{eq:fulllike}
    \begin{aligned}
    \ell_{full}(\bm{\theta}|\bm{l},\bm{\xi})=&\log \Bigg[\prod_{i=1}^N \prod_{g=1}^G \prod_{t=1}^{T_g} \Big[f_{\bm{L}_{i,g_t}|\bm{\xi}_{i,g_t}}(\bm{l}_{i,g_t}|\bm{\xi}_{i,g_t};m,\bm{\beta})\Big] \\
    &\times \prod_{i=1}^N \prod_{g=1}^G \Big[ f_{\bm{\xi}_{ig}}(\bm{\xi}_{i,g_1}, ..., \bm{\xi}_{i,g_t}, ..., \bm{\xi}_{i,g_{T_g}};M, p)\Big]\Bigg],
    \end{aligned}
\end{equation}
where the term $f_{\bm{L}_{i,g_t}|\bm{\xi}_{i,g_t}}(\bm{l}_{i,g_t}|\bm{\xi}_{i,g_t};m,\bm{\beta})$ is the conditional density of $\bm{L}_{i,g_t}$ given $\bm{\xi}_{i,g_t}$ and the term $f_{\bm{\xi}_g}(\bm{\xi}_{i,g_1}, ..., \bm{\xi}_{i,g_t}, ..., \bm{\xi}_{i,g_{T_g}};M, p)$ is the joint density function of $[\bm{\xi}_{i,g_1}, ..., \bm{\xi}_{i,g_t}, ..., \bm{\xi}_{i,g_{T_g}}]$. 

\citet[][Theorem 3.2.18]{muirhead2009aspects} gives the expression of $f_{\bm{L}_{i,g_t}|\bm{\xi}_{i,g_t}}(\bm{l}_{i,g_t}|\bm{\xi}_{i,g_t};m,\bm{\beta})$. The density function is difficult to compute due to the presence of an integral which is difficult to evaluate. Given the assumption that $m$ is assumed to be large to make $\mathbb{E}_{L_{it}^{(k)}|\xi_{it}^{(k)}}L_{it}^{(k)}$ dominated by $\lambda_{i,g_t}^{(k)}=\xi_{i,g_t}^{(k)}e^{\beta_{0,g_t}^{(k)}+\bm{X}_i\bm{\beta}_{g_t}^{(k)}}$, we give Lemma \ref{thm:thm0} borrowed from \citet[][Corollary 9.4.2]{mmbook1979}.
\begin{lemma}
\label{thm:thm0}
In Equation \ref{eq:bbb}, $\sqrt{m}\frac{L_{i,g_t}^{(k)}-\lambda_{i,g_t}^{(k)}}{ \sqrt{2}\lambda_{i,g_t}^{(k)} }$ converges in distribution to a standard normal distribution as $m\rightarrow\infty$, conditional on $\xi_{it}^{(k)}$.
\end{lemma}
Therefore, to simplify the computation, we approximate $f_{\bm{L}_{i,g_t}|\bm{\xi}_{i,g_t}}(\bm{l}_{i,g_t}|\bm{\xi}_{i,g_t};m,\bm{\beta})$ as a product of probability density functions of normal distributed random variables, expressed as
\begin{equation}
\label{eq:lll}
\begin{aligned}
    f_{\bm{L}_{i,g_t}|\bm{\xi}_{i,g_t}}(\bm{l}_{i,g_t}|\bm{\xi}_{i,g_t};m,\bm{\beta})\simeq \prod_{k=1}^3\phi\Bigg(l_{i,g_t}^{(k)};\lambda_{i,g_t}^{(k)},\sqrt{\frac{2}{m}}\lambda_{i,g_t}^{(k)} \Bigg),
\end{aligned} 
\end{equation}
where $\lambda_{i,g_t}^{(k)}=\xi_{i,g_t}^{(k)}e^{\beta_{0,g_t}^{(k)}+\bm{X}_i\bm{\beta}_{g_t}^{(k)}}$ and $\phi(l;\mu,\sigma)$ is the probability density function of a normal distribution with mean $\mu$ and standard error $\sigma$. In practice, this simplification reduces the computational cost significantly but preserves the accurate parameter estimation.

Next we handle the probability density function $f_{\bm{\xi}_{ig}}(\bm{\xi}_{i,g_1}, ..., \bm{\xi}_{i,g_t}, ..., \bm{\xi}_{i,g_{T_g}};M, p)$. In the following paragraphs, we derive the expression of the probability density function. First, Lemma \ref{thm:thm1} below allows us to handle only two density functions: $f_{\bm{\xi}_{i,g_1}}(\bm{\xi}_{i,g_1};M)$ and $f_{\bm{\xi}_{i,g_t}|\bm{\xi}_{i,g_{t-1}}}(\bm{\xi}_{i,g_t}|\bm{\xi}_{i,g_{t-1}};M, p)$.
\begin{lemma}
\label{thm:thm1}
$f_{\bm{\xi}_{ig}}(\bm{\xi}_{i,g_1}, ..., \bm{\xi}_{i,g_t}, ..., \bm{\xi}_{i,g_{T_g}};M, p)$ in Equation \ref{eq:fulllike} can be written as $$f_{\bm{\xi}_{ig}}(\bm{\xi}_{i,g_1}, ..., \bm{\xi}_{i,g_t}, ..., \bm{\xi}_{i,g_{T_g}};M, p)=f_{\bm{\xi}_{i,g_1}}({\xi}_{i,g_1};M)\prod_{t=2}^{T_g}f_{\bm{\xi}_{i,g_t}|\bm{\xi}_{g_{i,t-1}}}(\bm{\xi}_{i,g_t}|\bm{\xi}_{g_{i,t-1}};M, p),$$
where $f_{\bm{\xi}_{i,g_1}}(\bm{\xi}_{i,g_1};M)$ is the density function of $\bm{\xi}_{i,g_1}=[\xi_{i,g_1}^{(1)}, \xi_{i,g_1}^{(2)}, \xi_{i,g_1}^{(3)}]$ and $f_{\bm{\xi}_{i,g_t}|\bm{\xi}_{i,g_{t-1}}}(\bm{\xi}_{i,g_t}|\bm{\xi}_{i,g_{t-1}};M, p)$ is the conditional density of $\bm{\xi}_{i,g_t}=[\xi_{i,g_t}^{(1)}, \xi_{i,g_t}^{(2)}, \xi_{i,g_t}^{(3)}]$ given $\bm{\xi}_{i,g_{t-1}}=[\xi_{i,g_{t-1} }^{(1)}, \xi_{i,g_{t-1}}^{(2)}, \xi_{i,g_{t-1}}^{(3)}]$.
\end{lemma}
Because $\bm{\xi}_{i,g_1}\sim\mathcal{E}([1,1,1],M)$, \citet[][Corollary 3.2.19]{muirhead2009aspects} directly gives the expression of the probability density function $f_{\bm{\xi}_{i,g_1}}(\bm{\xi}_{i,g_1};M)$, expressed
\begin{equation}
\label{eq:111}
\begin{aligned}
    &f_{\bm{\xi}_{i,g_1}}(\bm{\xi}_{i,g_1};M)=\frac{\pi^{3^2/2}}{(\frac{2}{M})^{3M/2}\Gamma_3(\frac{3}{2})\Gamma_3(\frac{M}{2})}\prod_{k=1}^3[\xi_{i,g_t}^{(k)}]^{(M-3-1)/2}\prod_{k<q}^{3}[\xi_{i,g_t}^{(k)}-\xi_{i,g_t}^{(q)}]\exp\left(-\frac{M}{2}\sum_{k=1}^3\xi_{i,g_t}^{(k)}\right).
\end{aligned}
\end{equation}
Given \citet[][Equation 7]{james1960distribution} and \citet[][Equation 68]{james1964distributions}, we further derives the expression of $f_{\bm{\xi}_{i,g_t}|\bm{\xi}_{i,g_{t-1}}}(\bm{\xi}_{i,g_t}|\bm{\xi}_{i,g_{t-1}};M, p)$ (see Appendix \ref{sec:derivations}), expressed as
\begin{equation}
\label{eq:conditional}
\small
    \begin{aligned}
        f_{\bm{\xi}_{i,g_t}|\bm{\xi}_{i,g_{t-1}}}(\bm{\xi}_{i,g_t}|\bm{\xi}_{i,g_{t-1}};M, p)=&\frac{\pi^{3^2/2}}{(\frac{2}{M})^{3M/2}\Gamma_3(\frac{3}{2})\Gamma_3(\frac{M}{2})}\prod_{k=1}^3[\xi_{i,g_t}^{(k)}]^{(M-3-1)/2}\prod_{k<q}^{3}[\xi_{i,g_t}^{(k)}-\xi_{i,g_t}^{(q)}]\\
        &\times\exp\Bigg[-\frac{M}{2} \Big( \frac{p^2}{1-p^2}\sum_{k=1}^K\xi_{i,g_{t-1}}^{(k)} + \frac{1}{1-p^2}\sum_{k=1}^K\xi_{i,g_{t}}^{(k)}\Big) \Bigg]\times \big(\frac{1}{1-p^2}\big)^{\frac{3M}{2}}\\
        &\times \ _0F_1\Big(\frac{1}{2}M;\frac{M}{2}\frac{p}{1-p^2}[\xi_{i,g_{t-1}}^{(1)},\xi_{i,g_{t-1}}^{(2)},\xi_{i,g_{t-1}}^{(3)}],\frac{M}{2}\frac{p}{1-p^2}[\xi_{i,g_t}^{(1)},\xi_{i,g_t}^{(2)},\xi_{i,g_t}^{(3)}]\Big),
    \end{aligned}
\end{equation}
where $\ _0F_1(;)$ is a hypergeometric function of a matrix argument and its value can be numerically evaluated \citep{koev2006efficient}. When $p=0$, $\bm{\xi}_{i,g_t}$ is independent of $\bm{\xi}_{i,g_{t-1}}$. Furthermore, either $f_{\bm{\xi}_{i,g_t},\bm{\xi}_{i,g_{t-1}}}(\bm{\xi}_{i,g_t},\bm{\xi}_{i,g_{t-1}};M, p)=f_{\bm{\xi}_{i,g_t}|\bm{\xi}_{i,g_{t-1}}}(\bm{\xi}_{i,g_t}|\bm{\xi}_{i,g_{t-1}};M, p)f_{\bm{\xi}_{i,g_{t-1}}}(\bm{\xi}_{i,g_{t-1}};M)$ or $f_{\bm{\xi}_{i,g_t},\bm{\xi}_{i,g_{t-1}}}(\bm{\xi}_{i,g_t},\bm{\xi}_{i,g_{t-1}};M, p)=f_{\bm{\xi}_{i,g_{t-1}}|\bm{\xi}_{i,g_{t}}}(\bm{\xi}_{i,g_{t-1}}|\bm{\xi}_{i,g_{t}};M, p)f_{\bm{\xi}_{i,g_t}}(\bm{\xi}_{i,g_t};M)$ produces the joint density function of $\bm{\xi}_{i,g_t}$ and $\bm{\xi}_{i,g_{t-1}}$, denoted as $f_{\bm{\xi}_{i,g_t},\bm{\xi}_{i,g_{t-1}}}(\bm{\xi}_{i,g_t},\bm{\xi}_{i,g_{t-1}};M, p)$.

Putting Equations \ref{eq:lll}-\ref{eq:conditional} together, we have the explicit expression of the full log-likelihood $\ell_{full}(\bm{\theta}|\bm{l},\bm{\xi})$ (Equation \ref{eq:fulllike}). Standing on the above results, we next proceed the statistical inference of correlated eigenvalues. Our goal is to obtain the maximum likelihood estimators of $\bm{\theta}$. In particular, to answer the clinical question about covariate effect quantification, the maximum likelihood estimate of $\bm{\beta}_{g_t}^{(k)}$ quantifies the covariate effects, and its confidence interval describes the uncertainties of covariate effects. Therefore, we have to maximize the likelihood whose the latent variables $\bm{\xi}$ are integrated out. The log-likelihood $\ell(\bm{\theta}|\bm{l})$ whose latent variables are integrated out is expressed as
\begin{equation}
\begin{aligned}
    \ell(\bm{\theta}|\bm{l})=&\int \log \Bigg[\prod_{i=1}^N \prod_{g=1}^G \prod_{t=1}^{T_g} \Big[f_{\bm{L}_{i,g_t}|\bm{\xi}_{i,g_t}}(\bm{l}_{i,g_t}|\bm{\xi}_{i,g_t};m,\bm{\beta})\Big] \\
    &\times \prod_{i=1}^N \prod_{g=1}^G \Big[ f_{\bm{\xi}_g}(\bm{\xi}_{i,g_1}, ..., \bm{\xi}_{i,g_t}, ..., \bm{\xi}_{i,g_{T_g}};M, p)\Big]\Bigg]d\bm{\xi}.
    \end{aligned}
\end{equation}
The hurdle caused by the unobserved latent variables $\bm{\xi}$ can be easily overcome through the EM algorithm. The iterative scheme of the algorithm is given as follows. We start with the current parameter $\bm{\theta}^{(r)}$, thus the full likelihood at the current stage is $\ell_{Full}(\bm{\theta}^{(r)}|\bm{l},\bm{\xi})$.  We first compute the expected value of the likelihood function with respect to the current conditional distribution of $\bm{\xi}$ given the data $\bm{l}$ and the current parameter estimates $\bm{\theta}^{(r)}$, denoted as $Q(\bm{\theta}|\bm{\theta}^{(r)}) =\mathbb{E}_{\bm{\xi}|\bm{L},\bm{\theta}^{(r)}}\ell_{Full}(\bm{\theta}|\bm{l},\bm{\xi})$. This step is called E-step since we take the expectation.  Next we maximize the term $Q(\bm{\theta}|\bm{\theta}^{(r)})$ with respect to $\bm{\theta}$ to obtain the next estimates $\bm{\theta}^{(r+1)}$.  This step is called M-step because we maximize the function.  We iteratively repeat the two steps until convergence, in order to obtain the maximum likelihood estimator $\hat{\bm{\theta}}$ for the parameters $\bm{\theta}$. 

In the M-step, the function $Q(\bm{\theta}|\bm{\theta}^{(r)})$ can be feasibly maximized using the Quasi-Newton method with the the constraints ${\beta}_{dt}^{(1)}>{\beta}_{dt}^{(2)}>{\beta}_{dt}^{(3)}$ for $d=0,1,2,...,D$. In the E-step, the term $\mathbb{E}_{\bm{\xi}|\bm{L},\bm{\theta}^{(r)}}\ell_{Full}(\bm{\theta}|\bm{l},\bm{\xi})$ has no analytic expression. Thus, the Monte-Carlo EM based on importance sampling \citep[][Section 2.1]{levine2001implementations} is applied. We first generate $C$ Markov chain Monte-Carlo samples from the posterior distribution $\bm{\xi}|\bm{L},\bm{\theta}^{(0)}$, denoted as $\bm{\xi}^{(1)}, ..., \bm{\xi}^{(C)}$, where $\bm{\theta}^{(0)}$ is the initial values of the EM algorithm. For every step $r$, $Q(\bm{\theta}|\bm{\theta}^{(r)})$ is approximated as $Q(\bm{\theta}|\bm{\theta}^{(r)}) =\mathbb{E}_{\bm{\xi}|\bm{L},\bm{\theta}^{(r)}}\ell_{Full}(\bm{\theta}|\bm{l},\bm{\xi})\approx\frac{1}{C}\sum_{c=1}^C \omega_c \ell_{Full}(\bm{\theta}|\bm{l},\bm{\xi}^{(c)})$. The weight $\omega_c$ is defined as $\omega_c=\frac{\omega_c'}{\sum_{c=1}^C\omega_c'}$ where $\omega_c'=\exp\Big[\ell_{Full}(\bm{\theta}^{(r)}|\bm{l},\bm{\xi})-\ell_{Full}(\bm{\theta}^{(0)}|\bm{l},\bm{\xi})\Big]$. The correctness and relevant properties of the Monte-Carlo EM based on the importance sampling have already been provided in \citet{levine2001implementations}.

In clinical studies, the point and interval estimation of the coefficient ${\beta}_{d,g_t}^{(k)}$ are informative since it provides the covariate effects and the associated uncertainties, respectively. The confidence interval of parameters can be obtained by the parametric bootstrapping. We generate $B$ bootstrapping samples from Model \ref{eq:full1} with $\bm{\theta}=\hat{\bm{\theta}}$. We estimate ${\beta}_{d,g_t}^{(k)}$ for for each bootstrapping sample $b$, denoted as ${\beta}_{d,g_t}^{(k)*}(b)$, and we compute the bootstrapping difference as ${\delta}_{d,g_t}^{(k)*}(b)={\beta}_{d,g_t}^{(k)*}(b)-\widehat{\beta}_{d,g_t}^{(k)}$. The $1-\alpha$ confidence interval of ${\beta}_{d,g_t}^{(k)}$, is $[\widehat{\beta}_{d,g_t}^{(k)}-{\delta}_{d,g_t}^{(k)*}(b,1-\alpha/2),\hat{{\theta}}_i-{\delta}_{d,g_t}^{(k)*}(b,\alpha/2)]$, where ${\delta}_{d,g_t}^{(k)*}(b,1-\alpha/2)$ and ${\delta}_{d,g_t}^{(k)*}(b,\alpha/2)$ are the $1-\alpha/2$ and $\alpha/2$ quantiles of all the bootstrapping differences ${\delta}_{d,g_t}^{(k)*}(b)$ for $b=1, ..., B$. In the next section, we use simulation studies to compare out proposal to the other methods in terms of the point and interval estimation of the coefficient.

\section{Simulation Studies}
\label{sec:sim}
In this section, we carry out simulation studies to measure the performance of our proposal. We call our proposed model as \textit{auto-eigenvalues model}, and compare our method to a benchmark method. The benchmark model is a non-linear regression model treating the responses as normal distributed random variables. We refer this model as to the \textit{non-linear regression model}, expressed as $L_{i,g_t}^{(k)}=e^{\beta_{0t}^{(k)}+\bm{X}_i\bm{\beta}_{g_t}^{(k)}}+\epsilon_{i,g_t}^{(k)},\epsilon_{i,g_t}^{(k)}\sim\mathcal{N}(0,\sigma^2)$.

The two models have the similar interpretations of the coefficients $[\beta_{0t}^{(k)},\ {\bm{\beta}_{g_t}^{(k)}}^T]$, that are to control the mean of the random eigenvalues, but differ in basic statistical assumptions. To summary, the {auto-eigenvalues model} respects that the responses are random eigenvalues while the {non-linear regression model} does not. The {auto-eigenvalues model} additionally captures the correlation of the eigenvalues among voxels. More importantly, the interpretation of $e^{\beta_{0t}^{(k)}+\bm{X}_i\bm{\beta}_{t}^{(k)}}$ is actually different. Under the the {auto-eigenvalues model}, we have $\mathbb{E} L_{it}^{(k)}\simeq e^{\beta_{0t}^{(k)}+\bm{X}_i\bm{\beta}_{t}^{(k)}}$ as $M$ and $m$ are large. Under the {non-linear regression model}, there is no approximation involved, such that $\mathbb{E} L_{it}^{(k)}= e^{\beta_{0t}^{(k)}+\bm{X}_i\bm{\beta}_{t}^{(k)}}$. Therefore, we emphasize that the aim of the comparison is only to see whether we treat the responses as random eigenvalues or not impacts the coefficient estimation.

 The simulated data is generated from our proposed model. In each simulated data, we give the following settings. We have $G=3$ fibers and each fiber has $T_g=20$ voxels. The correlation parameter is set as $p=0.6$. The number of subjects is $N=10$. Since our model setup relies on large values of the degrees of freedom $m$ and $M$, we test parameter estimation performance given $M=m=20,30,40$, respectively. For each replication, we generate ${\beta}_{d,g_t}^{(1)}$ from a normal distribution with mean $0$ and standard error $0.1$, and we give ${\beta}_{d,g_t}^{(2)}={\beta}_{d,g_t}^{(1)}-0.3$ and ${\beta}_{d,g_t}^{(3)}={\beta}_{d,g_t}^{(1)}-0.6$.

\begin{figure}[ht!]
    \centering
    \includegraphics[width=0.95\textwidth]{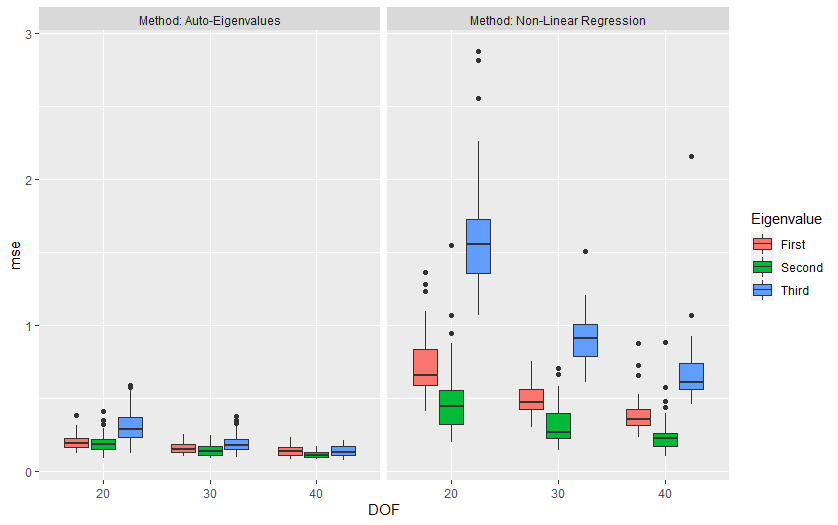}
    \caption{The figure gives the MSE of coefficients through boxplots. The left panel is for the auto-eigenvalues model, and the right panel is for the {non-linear regression model}. The x-axis is for the values of degrees of freedom. The y-axis is for the values of MSE.}
    \label{fig:MSE}
\end{figure}

We first evaluate the mean square error (MSE) of the coefficient parameter estimates. A smaller MSE indicates that the estimation is more accurate. The MSE is reported for each eigenvalue, defined as $\frac{1}{D\times\sum_{g}g_T}\sum_d\sum_g\sum_{g_t}({\beta}_{d,g_t}^{(k)}-{\beta}_{d,g_t}^{(k)})^2$. We use box-plots to visualize the MSEs of each replications (see Figure \ref{fig:MSE}). 
Regarding our proposed method, we find the trivial impact of the degrees of freedom on the parameter estimation of the coefficients, although the larger value of the degrees of freedom produces slightly more accurate results. This validates that our approximation in Equation \ref{eq:lll} is acceptable. Since the interpretations of the coefficients are not equal between the two models, the MSEs of the two methods are not very comparable in terms of their crude numbers. An important observation is that the three eigenvalues in the {non-linear regression} present heterogeneous MSEs, even if the values of the degrees of freedom are large. This should be caused by that the {non-linear regression model} treats the three random eigenvalues equally via ignoring the random process of eigenvalues. 

Furthermore, we compare the two methods in terms of the $95\%$ coverage probability. The $95\%$ coverage probability is defined as the probability that the true value is covered by the estimated $95\%$ confidence interval. The confidence interval of the {non-linear regression model} is also obtained via parametric bootstrapping. In Table \ref{tab:conf}, we summarize the coverage probabilities associated with each eigenvalue $k$, i.e., $[\beta_{0t}^{(k)},\ {\bm{\beta}_{g_t}^{(k)}}^T]$, averaging over voxels, coefficients, simulation replications. The {auto-eigenvalue model} presents reasonable uncertainties of parameter estimation. However, the {non-linear regression model} presents an an inflated confidence interval due to the model misspecification, where the parameter estimation of the coefficients has inflated uncertainties.

\begin{table}[ht!]
\caption{The table summarizes the $95\%$ coverage probabilities of the {auto-eigenvalue model} and the {non-linear regression model}.}
\label{tab:conf}
\centering
\begin{tabular}{c|cccc}
\hline\hline
\multirow{2}{*}{Method} & \multirow{2}{*}{Eigenvalue} & \multicolumn{3}{c}{Degrees of Freedom} \\ \cline{3-5} 
 &  & 20 & 30 & 40 \\ \hline\hline
\multirow{3}{*}{Auto-eigenvalues Model} & First & 90\% & 91\% & 92\% \\
 & Second & 97\% & 95\% & 94\% \\
 & Third & 99\% & 98\% & 97\% \\ \hline
\multirow{3}{*}{Non-linear Regression} & First & 100\% & 100\% & 100\% \\
 & Second & 100\% & 100\% & 100\% \\
 & Third & 100\% & 100\% & 100\% \\ \hline\hline
\end{tabular}
\end{table}

\section{Applications to IXI Dataset}
\label{sec:app}
\begin{figure}[ht!]
    \centering
    \includegraphics[width=0.9\textwidth]{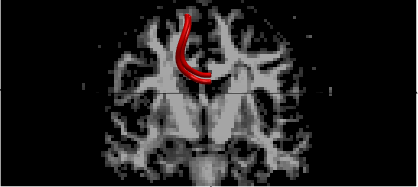}
    \caption{The figure illustrates the fibers the two fibers located at corpus callosum. The background is the Coronal plane of a human brain. The two red bundles are the fibers located in the region of interest.}
    \label{fig:real_fiber}
\end{figure}

We use the IXI data-set provided by the Center for the Developing Brain, Imperial College London, to demonstrate our proposed method. The data-set is made available at the website \url{https://brain-development.org/ixi-dataset/} under the CC BY-SA 3.0 license. From the data repository, we select the first $20$ healthy subjects collected from Hammersmith Hospital in London to demonstrate our methodological contribution and compare it to the benchmark method. We focus on the effects of body/weight index (BWI) on the two fibers located at the corpus callosum (see Figure \ref{fig:real_fiber}). The covariate of BWI is scaled into $[0,1]$.

We continue to compare our proposed {auto-eigenvalue model} to the alternative method {non-linear regression model}. To evaluate the covariate effects, we report the Z-socres of the coefficient estimators, which is defined as $\frac{\widehat{\beta}_{d,g_t}^{(k)}}{\sigma_{dt}^{(k)}}$, where $\sigma_{dt}^{(k)}$ is the standard error of the estimator $\widehat{\beta}_{d,g_t}^{(k)}$. A larger absolute value of z-score indicates that the corresponding covariate has more significant effect on $k$-th eigenvalue located at the voxel $g_t$.

\begin{figure}[ht!]
    \centering
    \includegraphics[width=0.85\textwidth]{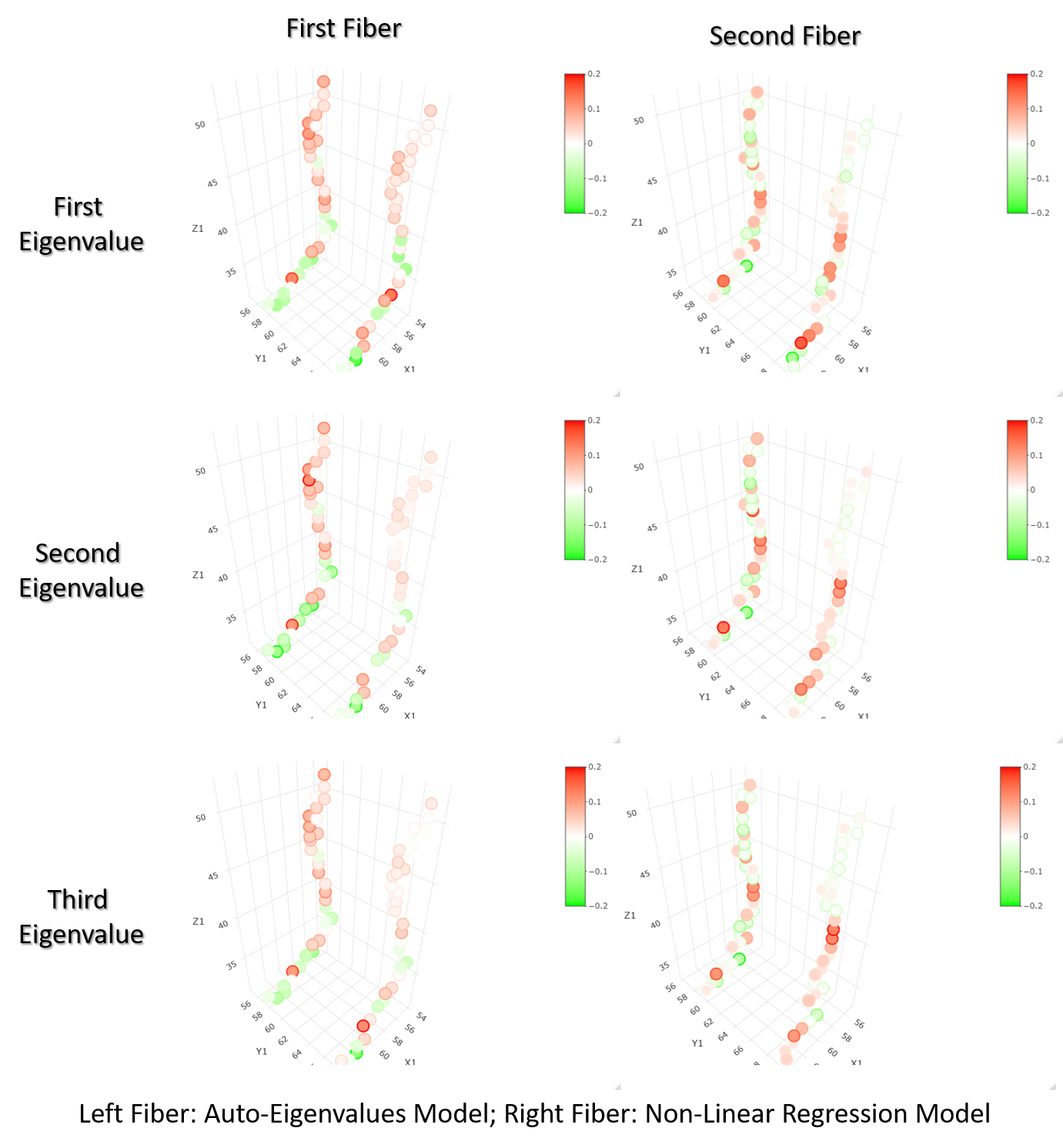}
    \caption{This figure gives the covariate effects of BWI. In each panel, the left fiber is the result of the auto-eigenvalues model and the right fiber is the result of the non-linear regression model. The fills of the points show the the values of z-scores.}
    \label{fig:BWI}
\end{figure}

The covariate effects of BWI are given in Figure \ref{fig:BWI}. First, we observe that the two models' z-scores are roughly consistent. It indicates that the two models will not produce too different scientific interpretations. However, our proposed method is powerful in finding regions (voxels) with covariate effects since z-scores' absolute values are relatively larger. This is consistent with our results in the simulation studies (Section \ref{sec:sim}), that is the {non-linear regression model} presents inflated statistical uncertainties. Although the real statistical underlying process of the diffusion tensors' eigenvalues is intractable, this may further endorse our major claim in this paper: the responses are encouraged to be treated as random eigenvalues. Using our proposed model based on the distribution of Wishart's eigenvalues is an option. Furthermore, the point estimate and $95\%$ confidence interval of $p$ are $0.452$ and $[0.445,0.493]$, implying the existence of auto-correlation among random eigenvalues. The point estimate of $m$ and $M$ are $31.956$ and $32.795$, respectively. The $95\%$ confidence intervals of $m$ and $M$ are $[30.256,33.656]$ and $[ 31.573,33.940]$, respectively. The large value of degrees of freedom implies that the values of degrees of freedom might be large enough to make the term $e^{\beta_{0t}^{(k)}+\bm{X}_i\bm{\beta}_{g_t}^{(k)}}$ dominate the mean of the random eigenvalue $\mathbb{E}L_{i,g_t}^{(k)}$.

\section{Conclusion and Future Works}
\label{sec:con}
This paper focuses on the statistical inference of the diffusion tensors' eigenvalues because it is more scientifically interpretable. We propose a model of random eigenvalues based on the distribution of Wishart's eigenvalues. Relying on the previous works i.e., \citet[][Chapter 3.2.5]{muirhead2009aspects}, \citet{james1964distributions}, and \citet[Chapter 9.4]{mmbook1979}, we are able to construct the hierarchical model of random eigenvalues and derive the relevant theoretical details. The estimation based on the Monte-Carlo EM algorithm is also proposed. We use simulation to study the impact on the parameter estimation if the responses are not treated as random eigenvalues. The real data analysis using IXI data also suggests that the eigenvalue responses of DTI are encouraged to be treated as random eigenvalues since our proposal produces more powerful results.

In this paper, we focus on eigenvalues and do not yet explore the statistical inference of eigenvectors. \citet[Theorem 3.4.2]{mmbook1979} gives the density function expression of a Wishart distributed matrix's eigenvectors. More studies on eigenvectors' statistical inference are encouraged, including the specification of the covariate effect and correlation effect, and computational methods. All of these can be considered as future works.

\bigskip
\begin{center}
{\large\bf SUPPLEMENTARY MATERIAL}
\end{center}

\appendix

\section{Theoretical Details}
\label{sec:proofs}
In this section, we offer theoretical details which are not fully elaborated in the main contexts.

\subsection{Density Function of Parameterized Wishart Distribution}
\label{sec:density}
Let $\bm{W}\sim\mathcal{W}_p(\bm{\Sigma},U)$. The probability density function of $\bm{W}$ is 
\begin{equation}
    {\displaystyle f_{\mathbf {W} }(\mathbf {w} )={\frac {|\mathbf {w} |^{(U-p-1)/2}e^{-\operatorname {tr} (U\mathbf {\Sigma} ^{-1}\mathbf {w} )/2}}{2^{\frac {Up}{2}}|{\mathbf {\Sigma}/U }|^{U/2}\Gamma _{p}({\frac {U}{2}})}}}.
\end{equation}

\subsection{Proof of Lemma \ref{thm:thm1} and Derivation of Equation \ref{eq:conditional}}
\label{sec:derivations}
To make our proof and derivation concise, we omit the subscript $i$ in the following. First, the joint density function of $(\bm{Z}_{j,g_1}, ..., \bm{Z}_{j,g_{T_g}})$ denoted as $f_{\bm{Z}_{j,g_1}, ..., \bm{Z}_{j,g_t}, ..., \bm{Z}_{j,g_{T_g}}}(\bm{z}_{j,g_1}, ..., \bm{z}_{j,g_t}, ..., \bm{z}_{j,g_{T_g}})$ can be decomposed as
\begin{equation}
    \begin{aligned}
        &f_{\bm{Z}_{j,g_1}, ..., \bm{Z}_{j,g_t}, ..., \bm{Z}_{j,g_{T_g}}}(\bm{z}_{j,g_1}, ..., \bm{z}_{j,g_t}, ..., \bm{z}_{j,g_{T_g}})\\
        &=f_{\bm{Z}_{j,g_1}}(\bm{z}_{j,g_1})\times f_{\bm{Z}_{j,g_2}|\bm{Z}_{j,g_1}}(\bm{z}_{j,g_2}|\bm{z}_{j,g_1})\times f_{\bm{Z}_{j,g_3}|\bm{Z}_{j,g_2},\bm{Z}_{j,g_1}}(\bm{z}_{j,g_3}|\bm{z}_{j,g_2},\bm{z}_{j,g_1})\times\\
        &\quad...\\
        &\quad\times f_{\bm{Z}_{j,g_T}|\bm{Z}_{j,g_{T-1}},...,\bm{Z}_{j,g_1}}(\bm{z}_{j,g_T}|\bm{z}_{j,g_{T-1}},...,\bm{z}_{j,g_1}).
    \end{aligned}
\end{equation}
Since $(\bm{Z}_{j,g_1}, ..., \bm{Z}_{j,g_{T_g}})$ follows a multivariate normal distribution with the covariance which is a Toeplitz matrix and the inverse covariance matrix is a tridiagonal matrix, the joint density function $f_{\bm{Z}_{j,g_1}, ..., \bm{Z}_{j,g_t}, ..., \bm{Z}_{j,g_{T_g}}}(\bm{z}_{j,g_1}, ..., \bm{z}_{j,g_t}, ..., \bm{z}_{j,g_{T_g}})$ can be simplified as 
\begin{equation}
    \begin{aligned}
        &f_{\bm{Z}_{j,g_1}, ..., \bm{Z}_{j,g_t}, ..., \bm{Z}_{j,g_{T_g}}}(\bm{z}_{j,g_1}, ..., \bm{z}_{j,g_t}, ..., \bm{z}_{j,g_{T_g}})\\
        &=f_{\bm{Z}_{j,g_1}}(\bm{z}_{j,g_1})\times f_{\bm{Z}_{j,g_2}|\bm{Z}_{j,g_1}}(\bm{z}_{j,g_2}|\bm{z}_{j,g_1})\times f_{\bm{Z}_{j,g_3}|\bm{Z}_{j,g_2}}(\bm{z}_{j,g_3}|\bm{z}_{j,g_2})\times...\times f_{\bm{Z}_{j,g_T}|\bm{Z}_{j,g_{T-1}}}(\bm{z}_{j,g_T}|\bm{z}_{j,g_{T-1}}).
    \end{aligned}
\end{equation}

This decomposition indicates that the terms $(\bm{Z}_{j,g_1},[\bm{Z}_{j,g_2}|\bm{Z}_{j,g_1}], [\bm{Z}_{j,g_3}|\bm{Z}_{j,g_2}], ..., [\bm{Z}_{j,g_T}|\bm{Z}_{j,g_{T-1}}])$ are mutually independent. Therefore, the joint density function of $\bm{\xi}_{i,g_1}, ..., \bm{\xi}_{i,g_t}, ..., \bm{\xi}_{i,g_{T_g}}$ can be written as 
$$f_{\bm{\xi}_g}(\bm{\xi}_{i,g_1}, ..., \bm{\xi}_{i,g_t}, ..., \bm{\xi}_{i,g_{T_g}};M, p)=f_{\bm{\xi}_{g_1}}({\xi}_{i,g_1};M)\prod_{t=2}^{T_g}f_{\bm{\xi}_{g_t}|\bm{\xi}_{g_{t-1}}}(\bm{\xi}_{g_t}|\{\bm{Z}_{j,g_{t-1}}:j=1:M\};M, p),$$
because ${\xi}_{i,g_t}$ are the eigenvalues of $\bm{Z}_{j,g_t}\bm{Z}^T_{j,g_t}/M$.

Next we want to show that $f_{\bm{\xi}_{g_t}|\{\bm{Z}_{j,g_{t-1}}:j=1:M\}}(\bm{\xi}_{g_t}|\{\bm{Z}_{j,g_{t-1}}:j=1:M\};M, p)=f_{\bm{\xi}_{i,g_t}|\bm{\xi}_{i,g_{t-1}}}(\bm{\xi}_{i,g_t}|\bm{\xi}_{i,g_{t-1}};M, p)$. The derivation largely relies on \citet[][Equation 7]{james1960distribution} and \citet[][Equation 68]{james1964distributions}. The derivation trick is borrowed from \citet{smith2007distribution} where the conditional distribution of eigenvalues of a complex Wishart matrix is derived.

First, we re-illustrate \citet[][Equation 7]{james1960distribution} and \citet[][Equation 68]{james1964distributions} by giving Theorem \ref{thm:result}.
\begin{theorem}
\label{thm:result}
Let $\bm{K}_j\stackrel{ind}{\sim}\mathcal{N}(\bm{\mu}/\sqrt{M},\bm{\Sigma}/M)$. The roots of $|\sum_{j=1}^M\bm{K}_j\bm{K}_j^T-\bm{W}\bm{\Sigma}/M|=0$, denoted as $\bm{W}=diag[W_1,W_2,W_3]$ follow a distribution with the density function expressed as
\begin{equation}
\begin{aligned}
        &f_{W_1,W_2,W_3}(w_1,w_2,w_3)=\exp\Big[-\frac{1}{2}(w_1+w_2+w_3)\Big]\ _0F_1\Big(\frac{1}{2}M;\frac{1}{4}[o_1,o_2,o_3],[w_1,w_2,w_3]\Big)\\
    &\times \frac{\pi^{3^2/2}}{(\frac{2}{M})^{3M/2}\Gamma_3(\frac{3}{2})\Gamma_3(\frac{M}{2})}\exp\Big[-\frac{1}{2}(o_1+o_2+o_3)\Big]\prod_{k=1}^3(w_k)^{(M-3-1)/2}\prod_{k<q}^3[w_k-w_q],
\end{aligned}
\end{equation}
where the roots of $|\bm{\mu}\bm{\mu}^T-\bm{O}\bm{\Sigma}/M|=0$ are $\bm{O}=diag[o_1,o_2,o_3]$.
\end{theorem}
We next apply this theorem to our problem based on the result of conditional normal distribution, i.e., $[\bm{Z}_{j,g_t}|\bm{Z}_{j,g_{t-1}}]\sim\mathcal{N}({p}\bm{Z}_{j,g_{t-1}},(1-{p}^2)\bm{I})$. By letting $\bm{K}_j$ in Theorem \ref{thm:result} be $[\bm{Z}_{j,g_t}|\bm{Z}_{j,g_{t-1}}]$, we have 
\begin{equation}
\begin{aligned}
        &f_{W_1,W_2,W_3}(w_1,w_2,w_3)=\exp\Big[-\frac{1}{2}(w_1+w_2+w_3)\Big]\ _0F_1\Big(\frac{1}{2}M;\frac{M{p}^2}{4(1-{p}^2)}[\xi_{i,g_{t-1}}^{(1)},\xi_{i,g_{t-1}}^{(2)},\xi_{i,g_{t-1}}^{(3)}],[w_1,w_2,w_3]\Big)\\
    &\times \frac{\pi^{3^2/2}}{(\frac{2}{M})^{3M/2}\Gamma_3(\frac{3}{2})\Gamma_3(\frac{M}{2})}\exp\Big[-\frac{M{p}^2}{2(1-{p}^2)}(\xi_{i,g_{t-1}}^{(1)}+\xi_{i,g_{t-1}}^{(2)}+\xi_{i,g_{t-1}}^{(3)})\Big]\prod_{k=1}^3(w_k)^{(M-3-1)/2}\prod_{k<q}^3[w_k-w_q],
\end{aligned}
\end{equation}
where $[W_1,W_2,W_3]=\frac{M}{(1-{p}^2)}[\xi_{i,g_{t-1}}^{(1)},\xi_{i,g_{t-1}}^{(2)},\xi_{i,g_{t-1}}^{(3)}]$. Via change of variables, we have
\begin{equation}
    \begin{aligned}
        f_{\bm{\xi}_{i,g_t}|\bm{\xi}_{i,g_{t-1}}}(\bm{\xi}_{i,g_t}|\bm{\xi}_{i,g_{t-1}};M, p)&=f_{W_1,W_2,W_3}(\frac{M}{(1-{p}^2)}\xi_{i,g_{t-1}}^{(1)},\frac{M}{(1-{p}^2)}\frac{M}{(1-{p}^2)}\xi_{i,g_{t-1}}^{(2)},\frac{M}{(1-{p}^2)}\xi_{i,g_{t-1}}^{(3)})\times|J|\\
        &\text{(Because $\ _0F_1(m;\bm{A},\bm{B})=\ _0F_1(m;\bm{A}/c,\bm{B}c)$, and $|J|=|\frac{M}{1-p^2}|^3$)}\\
        &=\frac{\pi^{3^2/2}}{(\frac{2}{M})^{3M/2}\Gamma_3(\frac{3}{2})\Gamma_3(\frac{M}{2})}\prod_{k=1}^3[\xi_{i,g_t}^{(k)}]^{(M-3-1)/2}\prod_{k<q}^{3}[\xi_{i,g_t}^{(k)}-\xi_{i,g_t}^{(q)}]\\
        &\times\exp\Bigg[-\frac{M}{2} \Big( \frac{p^2}{1-p^2}\sum_{k=1}^K\xi_{i,g_{t-1}}^{(k)} + \frac{1}{1-p^2}\sum_{k=1}^K\xi_{i,g_{t}}^{(k)}\Big) \Bigg]\times \big(\frac{1}{1-p^2}\big)^{\frac{3M}{2}}\\
        &\times \ _0F_1\Big(\frac{1}{2}M;\frac{M}{2}\frac{p}{1-p^2}[\xi_{i,g_{t-1}}^{(1)},\xi_{i,g_{t-1}}^{(2)},\xi_{i,g_{t-1}}^{(3)}],\frac{M}{2}\frac{p}{1-p^2}[\xi_{i,g_t}^{(1)},\xi_{i,g_t}^{(2)},\xi_{i,g_t}^{(3)}]\Big).
    \end{aligned}
\end{equation}

\section{Codes}
\label{sec:codes}
All the codes written in \code{MATLAB} are summarized in \code{AutoEigenCodes.zip}. The instructions of implementing the codes are provided \code{readme.txt}. Some example scripts are also given.

\end{document}